\documentclass[twocolumn,prb,nofootinbib]{revtex4}
\usepackage{graphicx}
\usepackage{color}

\begin{document}
\title{Persistent dynamic magnetic state in artificial honeycomb spin ice}

\author{J. Guo$^{1,\dagger}$}
\author{P. Ghosh$^{1,\dagger}$}
\author{D. Hill$^{1,\dagger}$}
\author{Y. Chen$^{2}$}
\author{L. Stingaciu$^{3}$}
\author{P. Zolnierczuk$^{3}$}
\author{C. A. Ullrich$^{1,*}$}
\author{D. K. Singh$^{1,*}$}

\affiliation{$^{1}$Department of Physics and Astronomy, University of Missouri, Columbia, MO}
\affiliation{$^{2}$Suzhou Institute of Nano-Tech and Nano-Bionics, Chinese Academy of Sciences, China}
\affiliation{$^{3}$Oak Ridge National Laboratory, Oak Ridge, TN 37831, USA}



\newcommand{\CU}{\color{red}}

\renewcommand{\thefootnote}{\fnsymbol{footnote}}

\begin{abstract}

\textbf{Topological magnetic charges, arising due to the non-vanishing magnetic flux on spin ice vertices, serve as the origin of magnetic monopoles that traverse the underlying lattice effortlessly. Unlike spin ice materials of atomic origin, the dynamic state in artificial honeycomb spin ice is conventionally described in terms of finite size domain wall kinetics that require magnetic field or current application. Contrary to this common understanding, here we show that thermally tunable artificial permalloy honeycomb lattice manifests a perpetual dynamic state due to self-propelled magnetic charge defect relaxation in the absence of any external tuning agent. Quantitative investigation of magnetic charge defect dynamics using neutron spin echo spectroscopy reveals sub-ns relaxation times that are comparable to monopole's relaxation in bulk spin ices. Most importantly, the kinetic process remains unabated at low temperature where thermal fluctuation is negligible. This suggests that dynamic phenomena in honeycomb spin ice are mediated by quasi-particle type entities, also confirmed by quantum Monte-Carlo simulations that replicate the kinetic behavior. Our research unveils a new `macroscopic' magnetic particle that shares many known traits of quantum particles, namely magnetic monopole and magnon.}

\end{abstract}

\maketitle

\footnotetext[2]{These authors contributed equally to the research.}
\footnotetext[1]{email: ullrichc@missouri.edu, singhdk@missouri.edu}

Magnetic structure and dynamics in a magnetic material are governed by the underlying energetics due to the interaction of magnetic moments with external and internal magnetic fields \cite{Coey,Gardner}. Unlike its classical counterpart, a quantum magnet manifests a persistent dynamic tendency associated with spin or magnetic moment relaxation.\cite{Gardner,Moessner2} A magnetic moment can be thought of as a pair of `$+$' and `$-$' quasi-particles, called magnetic charges, coupled via magnetic Coulomb interactions \cite{Sondhi}. This concept plays a dominant role in the temperature-dependent evolution of magnetic and physical properties in spin ice compounds \cite{Bramwell1,Bramwell2,Chern}. A direct demonstration of this effect is most discernible in two-dimensional artificial kagome ice on a honeycomb motif, where the possible local moment configurations (2-in and 1-out, 1-in and 2-out, and all-in or all-out) impart low ($Q$) and high ($3Q$) multiplicity magnetic charges on respective vertices \cite{SH,Nisoli,N,Parakkat}. Unequivocal evidence of magnetic charges has been obtained using magnetic force microscopy in honeycomb lattices made of permalloy (Ni$_{0.81}$Fe$_{0.19}$) and cobalt magnets \cite{Mengotti,Tanaka,Ladak,Shen}. However, they are not known to be dynamic in the absence of external tuning agents (magnetic, thermal, or current),\cite{SH,Nisoli,Shen,Heyderman} which is in strong contrast to the well-established dynamic properties of magnetic charges in analogous spin ice compounds that exhibit persistent or continuous relaxation at low enough temperature.\cite{Bramwell2,Jaubert,Broholm} Magnetic charges relax by emitting or absorbing net charge defects, $q_m$, of magnitude $2Q$, also termed effective monopoles.\cite{Sondhi,Gardner,SH,Mengotti,Zeissler}

\begin{figure*}
\includegraphics[width=\textwidth]{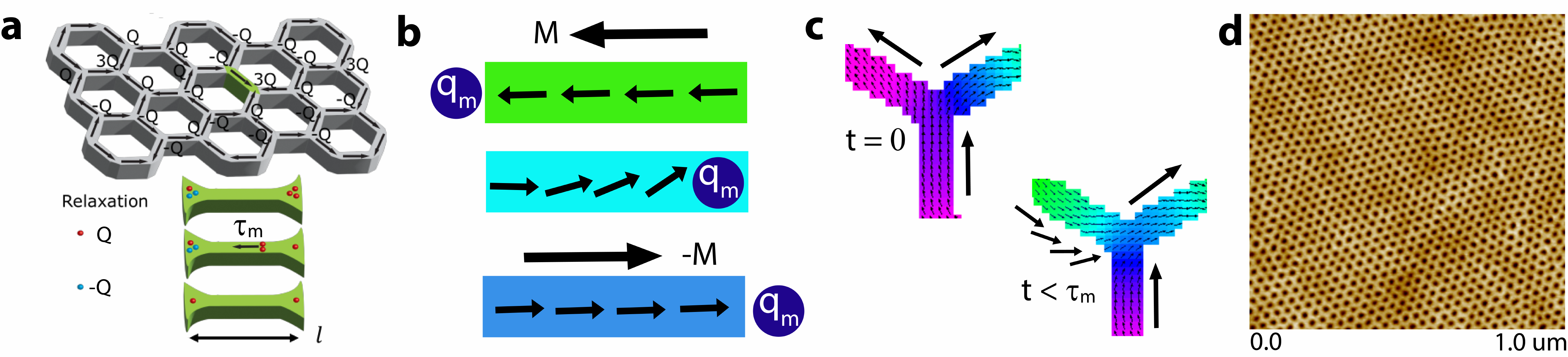}\vspace{-2mm}
\caption{\textbf{Self-propelled magnetic charge defect relaxation in an artificial honeycomb lattice made of ultra-small permalloy elements}. \textbf{a}, Schematic illustration of magnetic charge relaxation processes between honeycomb vertices. The charge defect ($q_m$) dynamics between nearest neighboring vertices corresponds to the reciprocal wave vector $q \sim 0.058$ \AA$^{-1}$. Magnetic charge relaxation occurs simultaneously in all three elements attached to a vertex. \textbf{b}, The dynamic state due to $q_{m}$ kinetics in a honeycomb element resembles magnon dynamics pattern. \textbf{c}, Time-dependent micromagnetic simulations depict magnetization pattern in a honeycomb element at two different instances, $t = 0$ and $t$ = 363 ps, of $q_m$ propagation. \textbf{d}, Atomic force micrograph of artificial honeycomb lattice, with typical element size of $\sim$ 11 nm (length)$\times$4 nm (width)$\times$6 nm (thickness).} \label{fig1}
\vspace{-6mm}
\end{figure*}

A key question is whether magnetic charge in artificial spin ice can be described by the same quantum mechanical treatment as in spin ice, where the Hamiltonian representing magnetic charge interaction is defined by Pauli matrices.\cite{Sondhi,Moessner2}. A direct comparison of magnetic charge relaxation times in spin ice and artificial magnetic honeycomb lattices can shed light on this fundamental problem. From the Coulombic physics perspective, a magnetic charge at the center of an artificial kagome lattice can be mapped onto the net magnetic flux along the bond direction on the parent honeycomb motif. \cite{Oleg, Moessner} The correspondence entails the quantum treatment of magnetic charge interactions, which has been used previously to deduce thermodynamic and magnetic properties of the statistical ensemble \cite{Oleg}. A charge defect's motion between vertices would flip the microscopic moment in the nanoscopic honeycomb element, manifesting a magnonic pattern, thereby altering the moment direction and the net charge on a given vertex, see Fig. \ref{fig1}a-c. The charge defect or magnonic charge, $q_m$, cannot pass through a high integer charge ($3Q$ or $-3Q$), which serves as a roadblock. At the same time, $\pm3Q$ charges have a high energy cost, which makes them unstable and triggers the dynamic process. A thermally tunable honeycomb lattice, made of ultra-small ($\sim$11 nm) nanoscopic permalloy element (see Fig. 1d),\cite{Summers} can host such dynamic event without the application of any external stimulus. It is known to depict a highly disordered ground state, comprised of both $\pm Q$ and $\pm 3Q$ charges, at low temperature.\cite{Yumnam} The ultra-small element size imparts thermally tunable energetics due to the modest inter-elemental magnetic dipolar interaction energy of $\sim$ 45 K. We use neutron spin echo (NSE) measurements to obtain direct evidence of self-propelled relaxation of magnonic charge $q_m$, as it provides accurate quantitative estimation of relaxation properties across a broad dynamic range.\cite{nse} Furthermore, NSE spectroscopy is carried out on two stacks of permalloy honeycomb samples with varying thicknesses of 6 nm and 8.5 nm that render a comprehensive outlook.

\begin{figure}
\includegraphics[width=\linewidth]{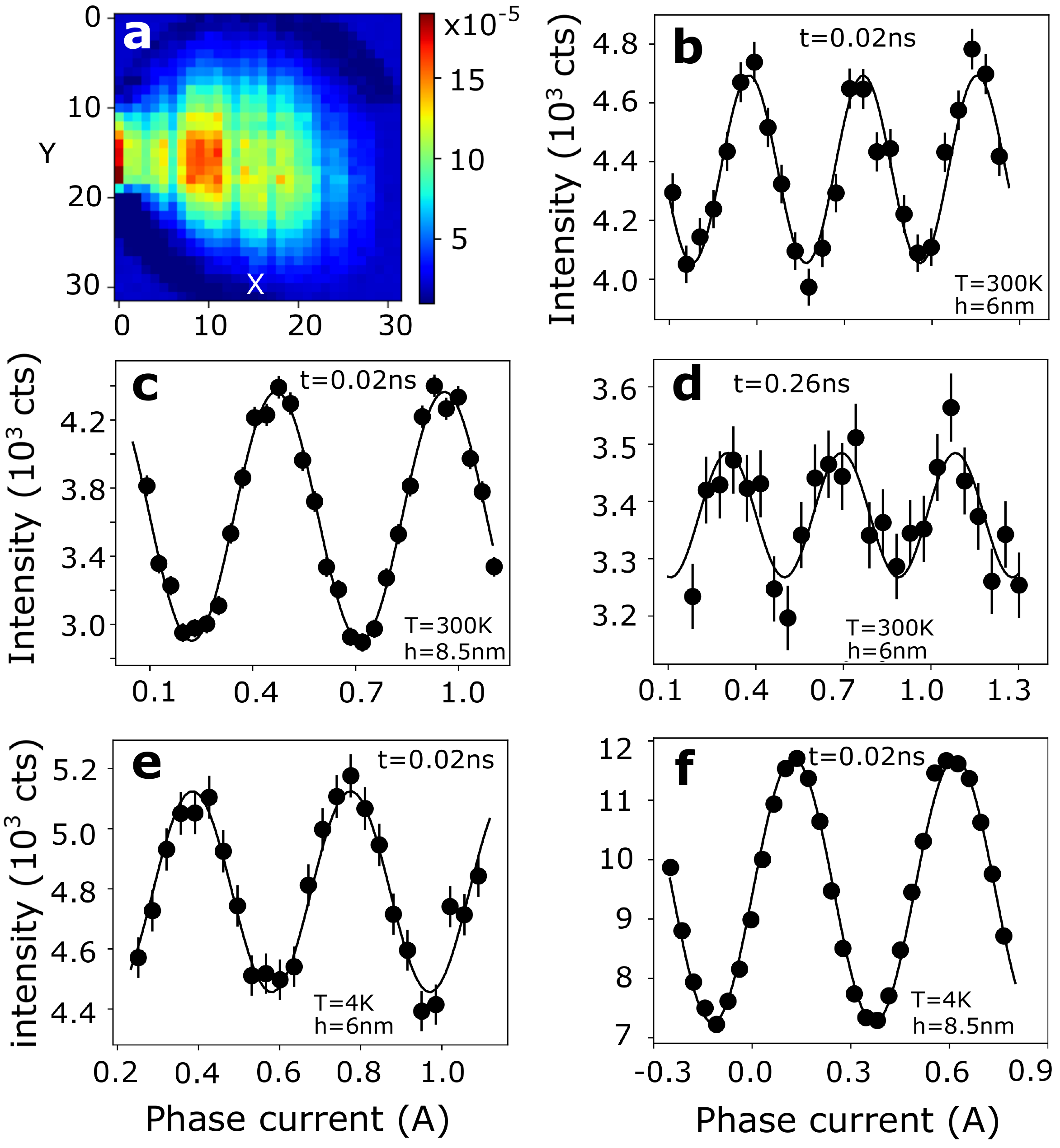}\vspace{-2mm}
\caption{\textbf{Neutron spin echo spectroscopy revealing dynamic properties in artificial honeycomb lattice without any external tuning parameter.} \textbf{a}, Color map of  NSE data on 6 nm thick honeycomb lattice with net neutron spin polarization $(+Z) - (-Z)$. Bright localized scattering pixels are identified with $q=0.058$ \AA$^{-1}$ and $q=0.029$ \AA$^{-1}$ (only partially visible near $x = 0$ due to geometrical limit of the instrument). NSE data is obtained on a large parallel stack of 125 (117) honeycomb samples of 6 nm (8.5 nm) thickness. \textbf{b,c}, Spin echo profile at $T = 300$ K in 6 nm and 8.5 nm thick samples, respectively, at $t = 0.02$ ns neutron Fourier time. Strong signal-to-background ratio is detected in the NSE signal.
\textbf{d}, The echo becomes weaker at higher Fourier time ($t$ = 0.26 ns), indicating the dominance of charge defect $q_m$ relaxation at lower time scale. \textbf{e,f}, Spin echo profile at $T = 4$ K in 6 nm and 8.5 nm thick samples, respectively, at $t = 0.02$ ns neutron Fourier time. Self-propelled charge dynamics prevails at low temperature.}
\label{fig2}
\vspace{-6mm}
\end{figure}

In Fig. \ref{fig2}a we show the color plot of NSE data obtained on a 6 nm thick honeycomb lattice with net neutron spin polarization along $+Z$ direction. No meaningful spectral weight is detected in the spin down neutron polarization (see Fig. S2-S3 in SM), \cite{supp} which confirms the magnetic nature of the signal. To ensure that the observed signal is magnetic in origin, NSE spectrometer is modified by removing the flipper before the sample (see Methods for detail). The figure exhibits very bright localized scattering pixels that are identified with $q = 0.058$ \AA$^{-1}$ and 0.029 \AA$^{-1}$. These $q$ values correspond to the typical relaxation length associated with the distance between the nearest and the next-nearest vertices, $2\pi/l$ and $2\pi/2l$, respectively, with $l = 11$ nm. This immediately suggests the quantized longitudinal nature of magnetic charge defect dynamics in artificial honeycomb lattices. NSE spectroscopy is a quasi-elastic measurement technique where the relaxation of magnetic specimen is decoded by measuring relative change in scattered neutron’s polarization via the change in the phase current at a given Fourier time (related to neutron precession). The localized excitations exhibit pronounced sinusoidal spin echo. Fig. \ref{fig2}b,c show the echo profiles obtained on honeycomb samples at room temperature at 0.02 ns Fourier time where a strong signal-to-background ratio is detected.

Typically, the echo intensity from a single wavelength $\lambda$ follows a cosine function, given by $I (\phi)$ = $A \cos(\phi \lambda) +B$, where the phase $\phi$ is directly related to the phase current $dJ$. However, if the neutron has a wavelength span of $\lambda_{\rm avg} \pm d\lambda$, then the signals from all different wavelengths add up (see SM). In this case, the total intensity follows the following functional relation,\cite{Piotr}
\begin{equation}
I(\phi)=
A \cos({\phi \lambda_{\rm avg}}) \frac{\sin(\phi d\lambda)}{\phi d\lambda} + B \:.
\end{equation}	
As the Fourier time associated to neutron precession increases, the error bar tends to become larger. Fig. \ref{fig2}d shows the spin echo at 0.26 ns Fourier time (see Fig. S4 in SM for other Fourier times). This trend hints at the fast relaxation of the magnetic charge quasi-particles, which is more prominent at a sub-ns time scale.

The sinusoidal nature of spin echo is quite evident at other temperatures as well. In Fig. \ref{fig2}e,f, we show the plot of characteristic spin echo profiles at low temperature in both thicknesses at $q= 0.058$ \AA$^{-1}$. Similar echo profiles are observed at $q=0.029$ \AA$^{-1}$. The strong spin echo at low temperature suggests the persistence of significant charge dynamics in the nanoscopic system. The origin of charge relaxation at low temperature of $T = 4$ K is unlikely to be thermal in nature considering that it costs $\mid$E$_{Q}$ - E$_{-Q}$$\mid$ ($\sim$ 76 K) to create a charge defect. Unlike other nanostructured magnets where the domain wall motion, typically ascribed to the dynamic property, is detectable in applied magnetic field or current only, the absence of any external tuning parameter at low temperature suggests the non-trivial nature of magnonic charge $q_m$ relaxation. In total, measurements were performed at six temperatures of $T$ = 4 K, 20 K, 50 K, 100 K, 200 K and 300 K (see Fig. S5 in SM). The quantitative estimation of relaxation times at different temperatures sheds light on the dynamic state of artificial honeycomb lattice.

The estimation of magnetic charge relaxation time is achieved by analyzing the intermediate scattering function $S(q, t)/S(q, 0)$ at different spin echo Fourier times. We show the plot of normalized intensity as a function of neutron Fourier time at $q=0.058$ \AA$^{-1}$ in the 6 nm thick sample in Fig. 3a. The normalization is achieved by dividing the observed oscillation amplitude by the maximum measurable amplitude (see SM), which is a common practice in magnetic systems with unsettling fluctuation to the lowest measurement temperature \cite{Piotr}. The normalized intensity reduces to the background level above the spin echo Fourier time of $\sim$ 0.5 ns, conforming to the most general signature of relaxation process in NSE measurements \cite{Ehlers,Piotr}. The quantitative determination of magnetic charge relaxation time involves
exponential fitting of the NSE scattering intensity as  $S(q,t)/S(q,0) = C e^{-t/\tau_m}$, where $C$ is a constant  \cite{Ehlers2}. The exponential function provides a good description of the relaxation mechanism of charge $q_m$.

Four important conclusions can be drawn from the estimated value of the relaxation time: first, the obtained value of $\tau_m$ = 20 ps at $T$ = 300 K suggests a highly active kinetic state in our permalloy honeycomb lattice, typically not observed in nanostructured materials where the domain wall motion is the primary mechanism behind the dynamic property \cite{Zhang}. Moreover, magnetic field or current application is necessary to trigger domain wall dynamics, which is not the case here: the dynamics state of charge $q_m$ is entirely self-propelled. Second, the relaxation rate of $q_m$ in honeycomb lattice is somewhat thickness independent. In Fig. 3b, we show $S(q, t)/S(q, 0)$ as a function of neutron Fourier time in the 8.5 nm thick lattice. The estimated $\tau_m = 49(10)$ ps is comparable to that found in the 6 nm thick lattice (within 2$\sigma$). Importantly, we do not observe a drastic difference between the charge dynamics in the thinner and the thicker lattice, which is a strong indication for its quasi-particle characteristic.

Third, the charge defect $q_m$ relaxes at the same high rate at low $T$ despite the absence of thermal fluctuation. As shown in Fig. 3c, the estimated relaxation time as a function of temperature suggests the persistence of charge dynamics at lower temperatures. Given the fact that the charge dynamics prevails in the absence of any external stimuli, the observed $T$-independence is quite significant. The conclusion is that the system lives in a persistent kinetic state due to the magnetic charge defect motion between parent lattice vertices, which is strong evidence for the quantum nature of magnetic charge quasi-particles.

Lastly, the $q_m$ relaxation rate in the 6 nm thick honeycomb lattice is comparable to that found in the bulk spin ice material, $\sim 5$ ps.\cite{Ehlers,Broholm} This suggests a correspondence between the dynamic behavior of magnetic charge defects and the monopolar dynamics in the atomistic spin ice. To our knowledge, this is the first observation of its kind, bridging the gap between the bulk spin ice material and the nanostructured spin ice system. Thus, the quasi-particle characteristic of the monopole can be suitably invoked in the artificial lattice. These conclusions are further supported by theoretical analysis, which we now briefly summarize.

In order to confirm the quantum nature of magnetic charge defect quasi-particles, we modeled the system both fully quantum mechanically and fully classically. Motivated by the simplistic behavior of magnetic charge relaxation, unchanged by temperature or system size, we suspect the system exhibits rapid renormalization group convergence to a minimal effective model of behavior, and as such we assume the system should be modeled effectively by a Hamiltonian with minimal features. To this end, for the quantum modeling we treat the magnetic monopoles at each vertex as a pseudospin 3/2 due to the quartet nature of these sites. These pseudospins are coupled minimally via nearest neighbor flip flop terms, i.e. $H_{nn} = \sum_{\langle ij \rangle} J_{h} S^+_i S^-_j$ where $S^+_i$ is the spin 3/2 representation of an SU(2) raising operator for the $i$th site and the sum is over nearest neighbors. We calculated the average relaxation time as a function of temperature for two cases, a 4-site model solved exactly, and a 200 site model with periodic boundary conditions time evolved via dynamic Monte Carlo.\cite{Adler}, Both cases reproduce the temperature independence of relaxation time for negligible barrier height compared to the large exchange constant, see Fig. \ref{fig4}a (see further detail in the SM).

\begin{figure}
\includegraphics[width=\linewidth]{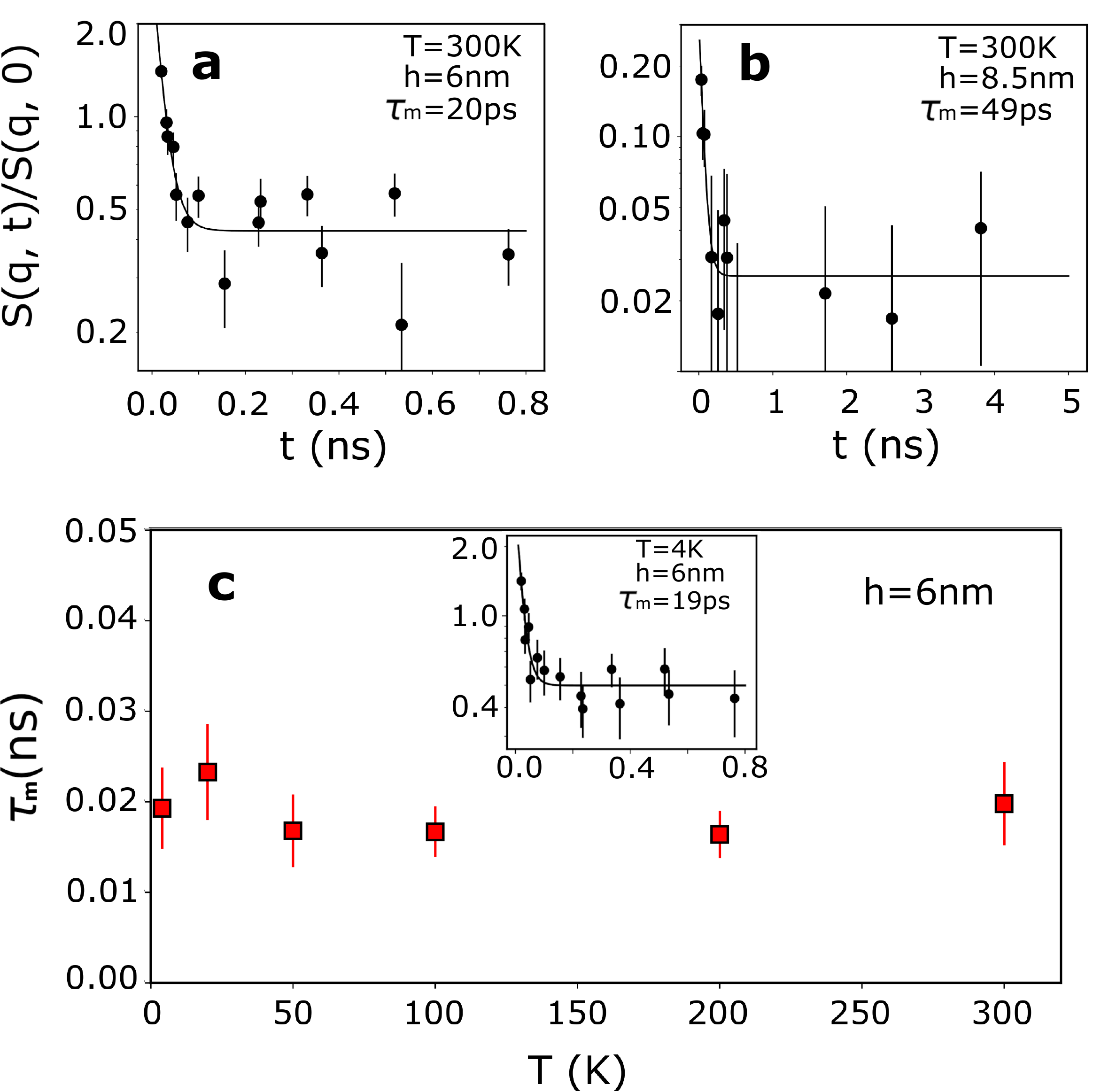}\vspace{-4mm}
\caption{\textbf{Temperature dependence of magnetic charge defect's relaxation time.} \textbf{a,b}, $S(q, t)/S(q, 0)$ versus neutron Fourier time at $T = 300$ K for the 6 nm sample (\textbf{a}) and the 8.5 nm sample (\textbf{b}).The magnetic charge relaxation time $\tau_m$ is obtained from an exponential fit of the data. \textbf{c}, $\tau_m$ versus $T$  in the 6 nm sample. Charge defect $q_m$ relaxes at $\tau$ $\sim$ 20 ps timescale at $T$ = 4 K without any external stimuli The plot also shows the temperature independence of $\tau_m$. Inset: same as (\textbf{a}), but for $T=4$ K.}
\label{fig3}
\vspace{-6mm}
\end{figure}

The main result of the quantum Monte-Carlo simulation is that a magnetic charge defect $q_m$ travels through the lattice element effortlessly as a result of negligible energy barrier. A similar observation was made in the case of magnetic monopole dynamics in atomic spin ice.\cite{Bramwell1} If the barrier height were comparable to the exchange constant $J$, then the relaxation time would increase significantly at low temperature; however, this is not seen in the experimental data. Similarly, if the kinetic behavior of $q_m$  were classical in origin, then the simulation would predict a monotonic temperature dependence of the relaxation time above the barrier height. 

For the classical modeling, we have treated each joint of the lattice as a classical spin obeying the Landau-Lifshitz-Gilbert equation with a stochastic field modeling temperature dependence \cite{Leliaert}. In this case the magnetic charge relaxation time was found to be temperature independent only up to a threshold temperature $T_c$, above which the relaxation time decreases monotonically with increasing temperature. This threshold occurs when the root-mean-square of the thermal field $\vec{B}_{th}$ is comparable in magnitude to the effective field term associated with shape anisotropy $\vec{B}_{an}$. With an estimated effective anisotropy (mostly shape anisotropy) of $|K_1| \sim 1 \times 10^4 \;\text{J/m}^3 $, we estimate the threshold temperature to be on the order of $T_c \sim 20$ K, see Fig. 4b. This is inconsistent with the experimental observations, leading us to conclude that the charge defect $q_m$ dynamics has indeed quantum nature.

\begin{figure}
\includegraphics[width=\linewidth]{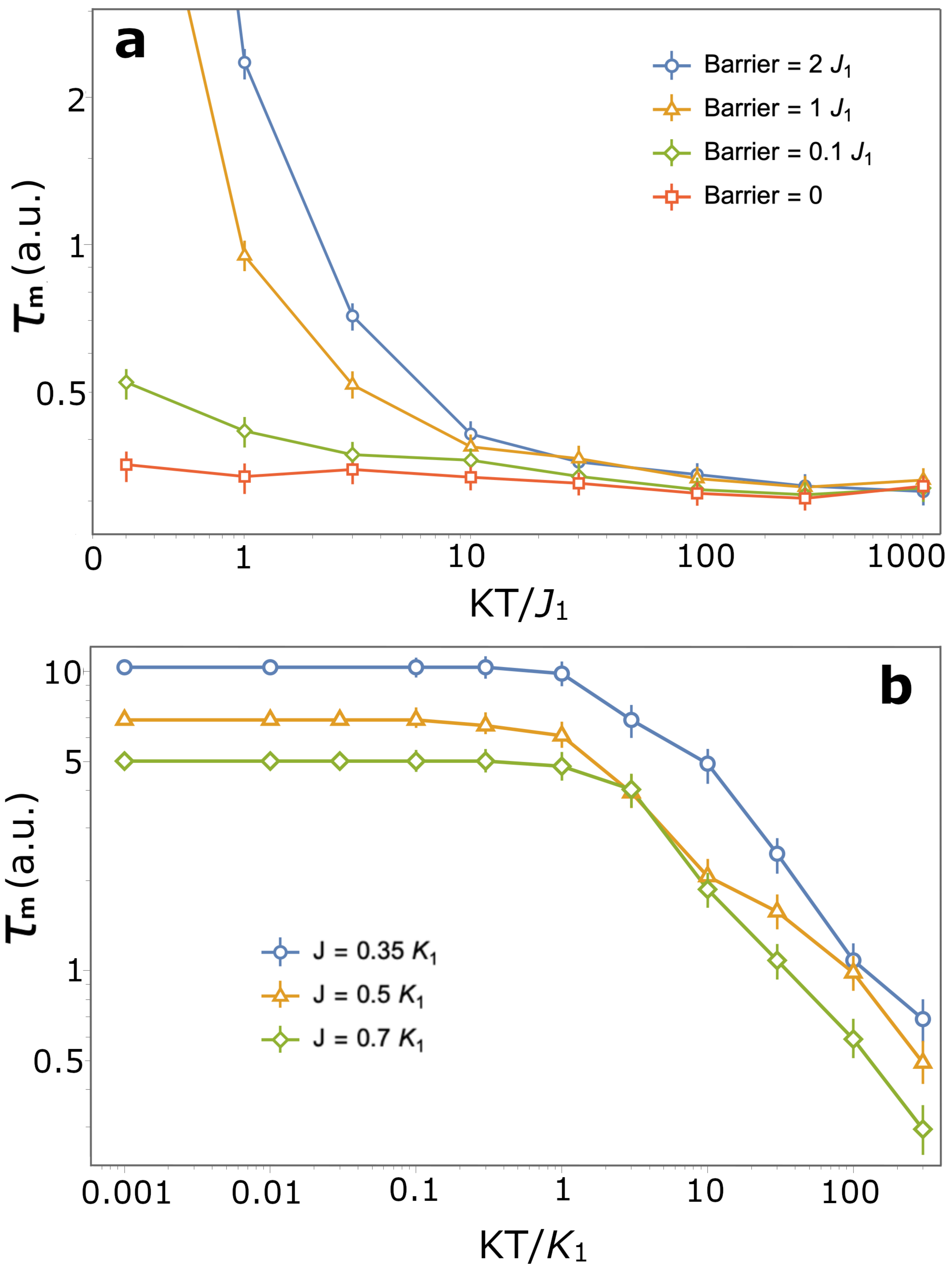}\vspace{-4mm}
\caption{\textbf{Quantum mechanical nature of $q_m$ relaxation process.} \textbf{a,b}, Temperature dependence of $\tau_m$ extracted from quantum (\textbf{a}) and classical (\textbf{b}) numerical simulations. Only the quantum Monte-Carlo simulations with negligible barrier height, compared to exchange constant, results in temperature independence of $\tau_m$, as found in experimental data. It suggests effortless relaxation of $q_m$, similar to quantum entity.}
\label{fig4}
\vspace{-6mm}
\end{figure}

This is the first time that a comprehensive study of magnetic charge relaxation processes in two-dimensional artificial permalloy honeycomb lattice reveals kinetic events with $\sim$ps temporal quantization. Previous efforts in determining the charge dynamics in artificial spin ice systems focused on the usage of ferromagnetic resonance spectroscopy and optical measurements such as Brillouin light scattering and optical cavity techniques \cite{Hoffman,Hastings}. However, these techniques have intrinsic temporal thresholds of 1 ns or higher, which limits the scope of sub-ns investigation of charge dynamics. Additionally, the samples used for dynamic study were created via electron-beam lithography, which results in large element size. In most cases, samples created in this way are athermal,\cite{Nisoli} thus significantly affecting the spatially quantized relaxation process between the vertices of the parent lattice. Our honeycomb sample with small inter-elemental energy, utilized in this study, provided an archetypal platform to extract the charge relaxation properties in the absence of thermal fluctuation. The ultra-small element size imparts a thermally tunable characteristic to the lattice. It is also relevant to mention that the macroscopic size honeycomb specimen, used in this study, tends to develop structural domains that vary in size from 400 nm to few $\mu$m. However, the NSE study of charge dynamics only involves the nearest and the next nearest neighboring vertices. Therefore, structural domains are not expected to affect the outcome.  The NSE technique, with a dynamic range extending to sub-ns timescales, is the most suitable experimental probe to study such delicate dynamic property.

The synergistic experimental and theoretical investigations have not only revealed the sub-20 ps relaxation time associated to magnonic charge dynamics but also elucidated its quantum mechanical characteristics. The thickness independence of the relaxation time $\tau$ gives further support to the  quasi-particle nature of the charge $q_m$. Despite the classical characteristic of magnetic charge, directly related to magnetic moment $M$ along the honeycomb element length ($L$) via $q = M/L$, the observed quantum mechanical nature of the charge defect dynamics is unprecedented. Unlike magnetic monopole or magnon particles in bulk magnetic materials of atomic origin, $q_m$ is expected to possess finite 'macroscopic' size. However, the kinetic behavior of $q_m$ manifests similarities with the known properties of magnetic monopoles, which is a strong indication of the particle-type character.

The new finding reported here can be utilized for the exploration of 'magnetricity' in artificial kagome ice, originally envisaged in the spin ice magnet.\cite{Bramwell2} Additionally, the quantum nature of magnonic charge can spur spintronic application via the indirect coupling with electric charge carriers.\cite{Chen} In recent times, experimental efforts are made to develop a new spintronic venue in artificial spin ice system. \cite{Kaffish} A dynamic system of thermally tunable magnetic honeycomb lattice renders a new research platform in this regard.

\textbf{Methods}

\textbf{Sample Fabrication}

Artificial honeycomb lattices were created using hierarchical top-down nanofabrication involving diblock copolymer templates [see details in Supplementary Material (SM) \cite{supp}]. An atomic force micrograph of the honeycomb lattice sample is shown in Fig. \ref{fig1}d.

\textbf{Neutron Spin Echo Measurements}

We performed NSE measurements on permalloy honeycomb lattice samples using an ultrahigh resolution (2 neV) neutron spectrometer SNS-NSE at beam line BL-15 of the Spallation Neutron Source, Oak Ridge National Laboratory. Time-of-flight experiments were performed using a neutron wavelength range of  3.5 -- 6.5 {\AA} and neutron Fourier times between 0.06 and 1 ns. NSE is a quasi-elastic technique where the relaxation of a magnetic specimen is decoded by measuring the relative polarization change of the scattered neutron via the change in the phase current at a given Fourier time (related to neutron precession). The experiment was carried out in a modified instrumental configuration of the NSE spectrometer where magnetism in the sample is used as a $\pi$ flipper to apply 180$^{\rm o}$ neutron spin inversion, instead of utilizing a flipper before the sample. Additional magnetic coils were installed to enable the XYZ neutron polarization analysis, see the schematic in Fig. S1 in SM. Such modifications ensure that the detected signal is magnetic in origin. NSE measurements were performed on two different parallel stacks of 125 and 117 samples of $20 \times 20$ mm surface size and 6 nm and 8.5 nm thicknesses, respectively, to obtain good signal-to-background ratio. The sample stacks were loaded in the custom-made sample cell, which was mounted in a close cycle refrigerator with 4 K base temperature. NSE data was collected for 8 hrs on the average at each temperature and neutron Fourier time.

\textbf{Acknowledgements}

We thank Valeria Lauter and George Yumnam for helping us understand the magnetic charge distribution on honeycomb vertices, and Antonio Faraone for helpful discussion on the use of NSE technique in probing magnetic materials. This work was supported by the U.S. Department of Energy, Office of Science, Basic Energy Sciences under Awards No. DE-SC0014461 (DKS) and  DE-SC0019109 (CAU). This work utilized the facilities supported by the Office of Basic Energy Sciences, US Department of Energy.

\textbf{Authors Contributions}

D.K.S and C.A.U. jointly led the research. D.K.S. envisaged the research idea and supervised every aspect of the experimental research. C.A.U supervised theoretical research. Samples were synthesized by J.G and P.G. Neutron spin echo measurements were carried out by J.G., Y.C., L.S., P.Z.  Analysis was carried out by J.G., L.S., P.Z. and D.K.S. Theoretical
calculations were carried out by D.H. and C.A.U. The paper was written by D.K.S. and C.A.U. with input of  all co-authors.

\textbf{Additional Information}

Supplementary Information is available for this paper.
Authors declare no competing financial interests. Correspondence and requests for materials should be addressed to D.K.S.

\clearpage
\end{document}


\title{Supplemental material for:\\Persistent dynamic magnetic state in artificial honeycomb spin ice}
\author{J. Guo$^{1,\dagger}$}
\author{P. Ghosh$^{1,\dagger}$}
\author{D. Hill$^{1,\dagger}$}
\author{Y. Chen$^{2}$}
\author{L. Stingaciu$^{3}$}
\author{P. Zolnierczuk$^{3}$}
\author{C. A. Ullrich$^{1,*}$}
\author{D. K. Singh$^{1,*}$}

\affiliation{$^{1}$Department of Physics and Astronomy, University of Missouri, Columbia, MO}
\affiliation{$^{2}$Suzhou Institute of Nano-Tech and Nano-Bionics, Chinese Academy of Sciences, China}
\affiliation{$^{3}$Oak Ridge National Laboratory, Oak Ridge, TN 37831, USA}
\affiliation{$^{*}$email: singhdk@missouri.edu, ullrichca@missouri.edu}
\affiliation{$^{\dagger}$These authors contribute equally to this work}

\maketitle

\section{Experimental details}

\subsection{Nanofabrication of the artificial honeycomb lattice of permalloy}

The fabrication of the artificial honeycomb lattice involves the synthesis of a porous hexagonal template on top of a silicon substrate, calibrated reactive ion etching to transfer the hexagonal pattern to the underlying silicon substrate and deposition of permalloy on top of the uniformly rotating substrate in a near-parallel configuration ($\sim$ 3$^{\circ}$) to achieve the 2D character of the system. The porous hexagonal template fabrication process utilizes diblock copolymer polystyrene (PS)-b-poly-4-vinly pyridine (P4VP) of molecular weight 29K Dalton and volume fractions 70\% PS and 30\% P4VP, which can self-assemble into hexagonal cylindrical structure of P4VP in the matrix of PS under the right condition. A PS-P4VP copolymer solution of mass fraction 0.6\% in toluene was spin-coated on a polished silicon wafer at around 2300 rpm for 30 seconds, followed by solvent vapor annealing at $\sim 25^{\circ}$C for 12 hours. A mixture of toluene/THF (20/80, volume fraction) was used for the solvent vapor annealing. This process results in the self-assembly of P4VP cylinders in a hexagonal pattern in a PS matrix.  Submerging the sample in ethanol for 20 minutes releases the P4VP cylinders from the PS matrix, leaving a hexagonal porous template with an average hole center-to-center distance around 31 nm. The diblock template is used as a mask to transfer the topographical pattern to the underlying silicon substrate using reactive ion etching with CF$_4$ gas. The top surface of the etched silicon substrate resembles a honeycomb lattice pattern, which is exploited to create a magnetic honeycomb lattice by depositing permalloy (Ni$_{0.81}$Fe$_{0.19}$) in a near-parallel configuration using E-beam physical vapor deposition. Samples were uniformed rotated to achieve uniformity of the deposition. This allowed evaporated permalloy to coat the top surface of the etched silicon substrate only and producing magnetic honeycomb lattice with a typical element size of about 11 nm (length) $\times$ 4 nm (width) and controllable thickness. A typical atomic force micrograph of the honeycomb lattice is shown in Fig. 1d. Samples of thickness 6 nm and 8.5 nm were used in this work.

\subsection{Time-of-flight Neutron Spin Echo instrument}

The Neutron Spin Echo (NSE) measurements were conducted on an ultrahigh resolution (2 neV corresponding to $\sim$ 300 ns Fourier time) neutron spectrometer at beam line BL--15 of the Spallation Neutron Source, Oak Ridge National Laboratory. A 3 {\AA} wavelength bandwidth of neutrons (3.5--6.5 \AA) was used in the experiment, and the scattered neutrons were collected with a 30 cm $\times$ 30 cm position-sensitive $^3$He detector 3.9 m away from the sample. The detector has 32 $\times$ 32 X, Y pixels laterally to keep track of the scattering angles of the detected neutrons. It also has 42 time of flight (ToF) channels (tbins) that label the time frame of the detected neutron which encodes the neutron's wavelength. Thus, the data collected by the detector is a 32 $\times$ 32 $\times$ 42 array. By carefully grouping the X, Y pixels and tbins, one can extract echo signals at different Fourier times $t$ [see Eq. (\ref{eq:fouriertime}) below for the definition] and $q$ values from a single measurement. In this experiment, echo signals at multiple temperatures and nominal Fourier times (the Fourier time associated with the neutron of the maximum wavelength) were measured.

\begin{figure}
    \includegraphics[width=0.8\textwidth]{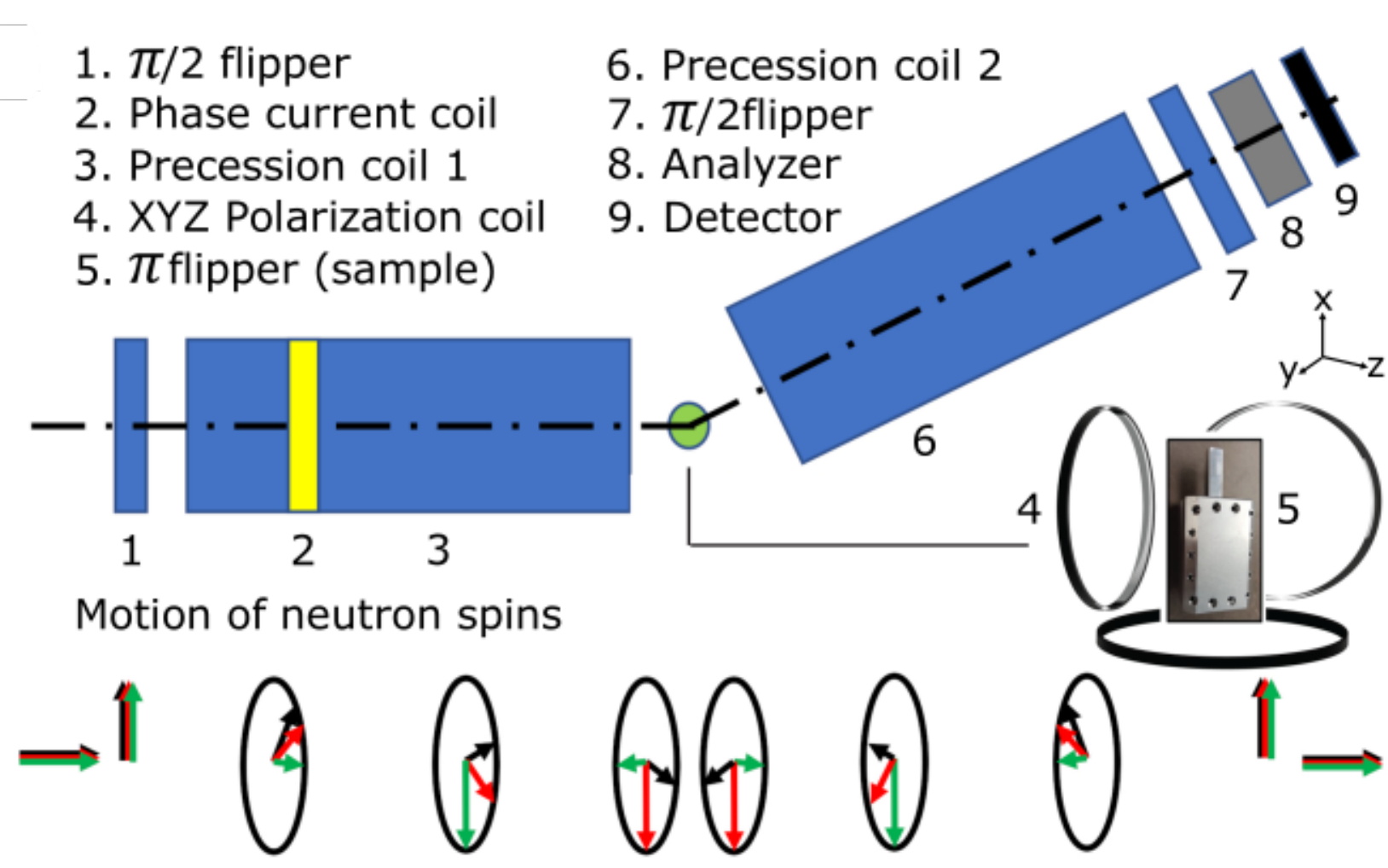}
    \caption{Design of modified NSE instrument, utilized for the magnetic charge relaxation experiment. Inset shows the stack of samples sealed in an aluminum can, which serves as a $\pi$ flipper. Additional magnetic coils are installed to enable the XYZ polarization analysis. }
    \label{fig:supp_instrument}
\end{figure}

Unlike in conventional NSE, where a $\pi$ flipper is installed before or after the sample to flip the neutron's polarization, a magnetic sample itself does a $\pi$ flip of the neutron's polarization due to the nature of the interactions between magnetic moments and neutron spins. In this experiment, the $\pi$ flipper was removed and additional magnetic coils were installed for three-dimensional polarization analysis. To obtain good signal-to-background ratio, a stack of 125 (117) samples of about 20$\times$20 mm$^2$ was loaded in a custom-made aluminum sample container. The sample container was inserted into a close cycle refrigerator with a base temperature of 4 K. A schematic diagram of the NSE instrument is presented in Fig. \ref{fig:supp_instrument}.

\subsection{Analysis of the Neutron Spin Echo signal}

NSE experiments probe the intermediate scattering function of the sample studied, given by
\begin{equation}
    S(q,t)=\int{\cos(\omega t)S(q,\omega)d\omega},
\end{equation}
where $S(Q,\omega)$ is the scattering function of the sample and $t$ is the Fourier time, defined as
\begin{equation}
    t=J\lambda^3\frac{\gamma_nm_n^2}{2\pi h^2}, \label{eq:fouriertime}
\end{equation}
with $\gamma_n$, $m_n$ and $\lambda$ being the neutron's gyromagnetic ratio, mass and wavelength, and $h$ denotes Planck's constant. $J=\int{|B|dl}$ is the magnetic field integral along the neutron's path through the precession coil \cite{piotr2019}. The field integrals are designed to be the same in both precession coils before and after the sample such that the precession phase acquired by the neutron in the first coil will be exactly recovered at the end of the second coil, given that the neutron does not change its velocity while interacting with the sample. For quasi-elastic scattering, the neutron gains a net phase at the end of the second coil before it reaches the analyzer. By systematically stepping through an additional phase current (convert to an additional field integral) in either the first or the second coil, a cosine modulation of the sample's intermediate scattering function is realized, which is the typical raw data, echo signal, one would analyze in an NSE experiment.

The echo signal intensity for a given neutron wavelength $\lambda$ has a cosine function shape given by
\begin{equation}
    I(\phi) = A\cos(\phi \lambda) + B,
\end{equation}
where $\phi=dJ \gamma_nm_n/h$ and $dJ$ is the phase asymmetry between the two precession coils introduced via the scanning phase current. When signals from neutrons of a certain wavelength span $\lambda_{\rm avg}\pm d\lambda$ are added up, the total intensity resembles a cosine function modulated by an envelope function
\begin{equation}
    I(\phi)=A\cos(\phi \lambda_{\rm avg})\frac{\sin(\phi d\lambda)}{\phi d\lambda}+B.
\end{equation}

\begin{figure}
    \includegraphics[width=\textwidth]{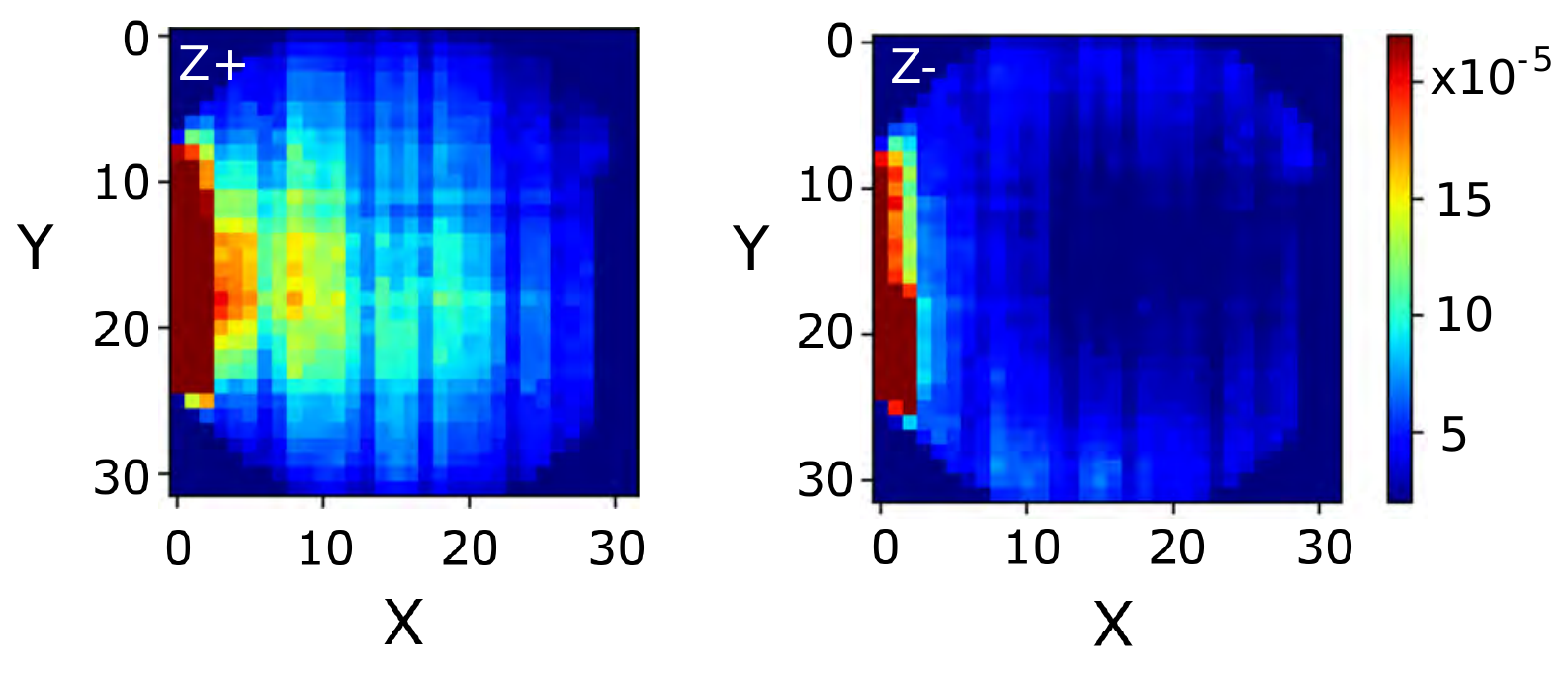}
    \caption{Color detector image of the 6nm honeycomb lattice at room temperature with neutron's polarization Z$+$ and Z$-$. The color scale on the right shows the detected neutron intensity normalized by the proton charge. With the same color scale, while no meaningful spectral weight is detected in the Z$-$ polarization, in the Z$+$ polarization case, clear spectral weight is observed in the detector region corresponding to $q\sim 0.058$ \AA$^{-1}$ and $q\sim 0.029$ \AA$^{-1}$. The intense signal at the left edge of the detector is the direct beam as the instrument hits its geometrical constraint limit. Fig. 2a in the main text is obtained by subtracting the Z$-$ intensities from the Z$+$ intensities.}
    \label{fig:suppl_detector_image}
\end{figure}

\begin{figure}
    \includegraphics[width=\textwidth]{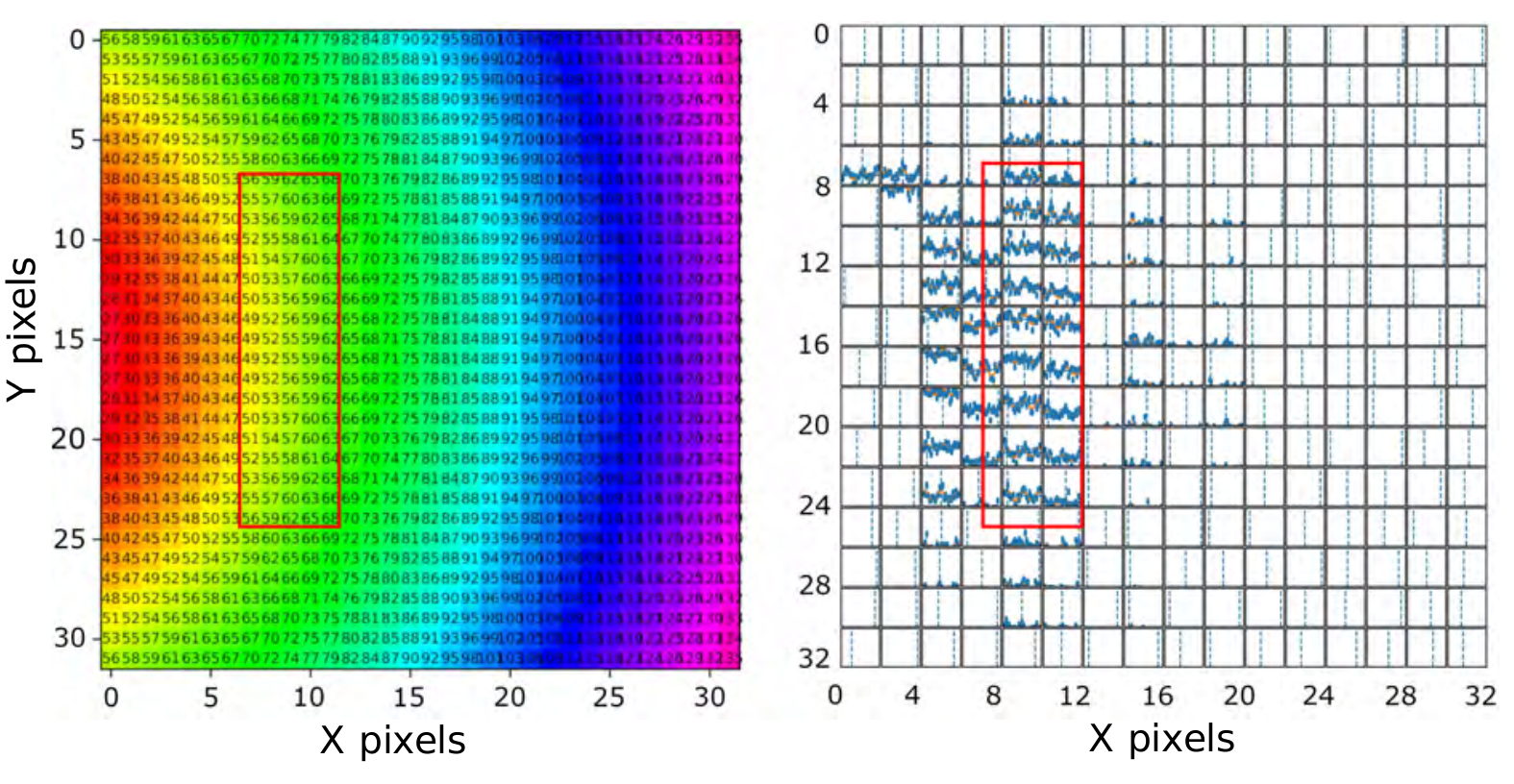}
    \caption{The $q$ distribution and echo signals of the 6 nm honeycomb lattice measured at T = 4 K on the detector. Left panel shows the flux-weighted $q$ distribution across the whole detector in ToF tbin 17, corresponding to a Fourier time of 0.02 ns, the numbers in the individual pixels denote the $q$ value in the unit of 0.001 \AA$^{-1}$. Right panel shows the echo signals in ToF tbin 17 on the detector. Every 2 pixels along X and Y directions are grouped for the clearness of the demonstration. The intense direct beam signals on the left edge of the detector are not shown. This image together with the color detector image in Fig. 2a in the main text guided the grouping of detector X, Y pixels in data analysis.}
    \label{fig:suppl_q_map}
\end{figure}

\subsection{Neutron Spin Echo raw data treatment}

The echo intensities were obtained by adding up signals in X, Y detector pixels and then summing over different ToF tbins while correcting the phase shifts among the tbins before the summation process. Grouping of the X, Y detector pixels were guided by the false color detector image (Fig. 2a in the main text) as well as the detector pixel image (Fig. \ref{fig:suppl_q_map}) to center around the intense scattering signal. The echo profiles  of the 8.5 nm honeycomb lattice shown in Fig. 2c,f of the main text were obtained by summing over detector pixels over the range of X = 8 to 10, Y = 10 to 21 in ToF tbin = 4, with a Fourier time $\sim$ 0.02 ns.  In order to make direct comparisons between the 6 nm honeycomb lattice and the 8.5 nm honeycomb lattice, echo profiles of the 6 nm honeycomb lattice (Fig. 2b,d,e of the main text) with the same Fourier time were obtained by summing over detector pixels over the range of X = 7 to 11, Y = 7 to 24 in ToF tbin = 17.

For the calculation of the intermediate scattering function (Fig. 3a,b of the main text), ToF tbins = 10 to 19, 20 to 29, 30 to 39 were explored. The detector pixel groupings used in the calculation of the intermediate scattering functions at each temperature are identical to that used for the echo profiles. Note that a slightly larger detector area is selected for the data analysis of the 6 nm honeycomb lattice. This is because the detected scattering signal from the 6 nm honeycomb lattice is broader than that from the 8.5 nm honeycomb lattice possibly due to a larger variation of the honeycomb element length scale.

\subsection{Calculation of the intermediate scattering function}

The goal of every NSE experiment is to obtain the intermediate scattering function $S(q,t)/S(q,0)$ as a function of the Fourier time $t$, which can be calculated from the fitted echo amplitude. For non-magnetic samples,
\begin{equation}
\frac{S(q,t)}{S(q,0)}=\frac{2A}{U-D}
\end{equation}
where $A$ denotes the fitted amplitude of the echo signal, $U$ and $D$ denote the spin up (non-$\pi$ flipping) and spin down ($\pi$ flipping) measurement of direct scattering without precession. $U-D$ measures the maximum obtainable echo amplitude. For magnetic samples, half of the magnetic scattering intensity $M/2$ is used for normalization \cite{pappas2006}. Thus 6 additional polarization measurements along $x$, $y$ and $z$ directions were performed at each temperature. $M$ and $U, D$ are calculated as
\begin{equation}
    M=2(z_{up}-z_{dn})-[(x_{up}-x_{dn})+(y_{up}-y_{dn})],
\end{equation}
\begin{equation}
        U=\frac{x_{up}+y_{up}+z_{up}}{3}, D=\frac{x_{dn}+y_{dn}+z_{dn}}{3}.
\end{equation}
The intermediate function is then determined as
\begin{equation}
    \frac{S(q,t)}{S(q,0)}=\frac{4A}{M},
\end{equation}
which is comparable with $66A/(U-D)$ in this experiment. In the calculation of the intermediate scattering function at a given Fourier time, the same X, Y pixels and ToF channels selection is used for the determination of both the echo amplitude and the normalization factor.

\begin{figure}
    \includegraphics[width=\textwidth]{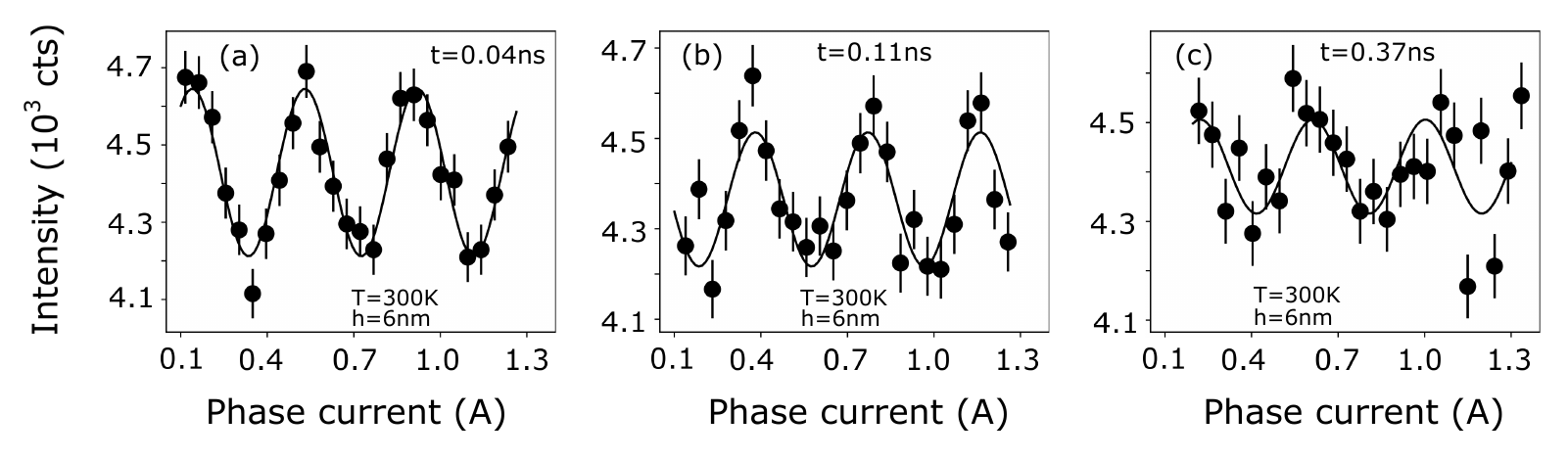}
    \caption{Echo profiles of the 6 nm honeycomb lattice at different Fourier times measured at room temperature, related to Fig. \ref{fig:suppl_q_map}. The Fourier time of the echoes are labeled on the graphs. Clearly, as the Fourier time increases, the error bars become large and the echoes exhibit smaller amplitudes.}
    \label{fig:echoes}
\end{figure}

\begin{figure}
    \includegraphics[width=\textwidth]{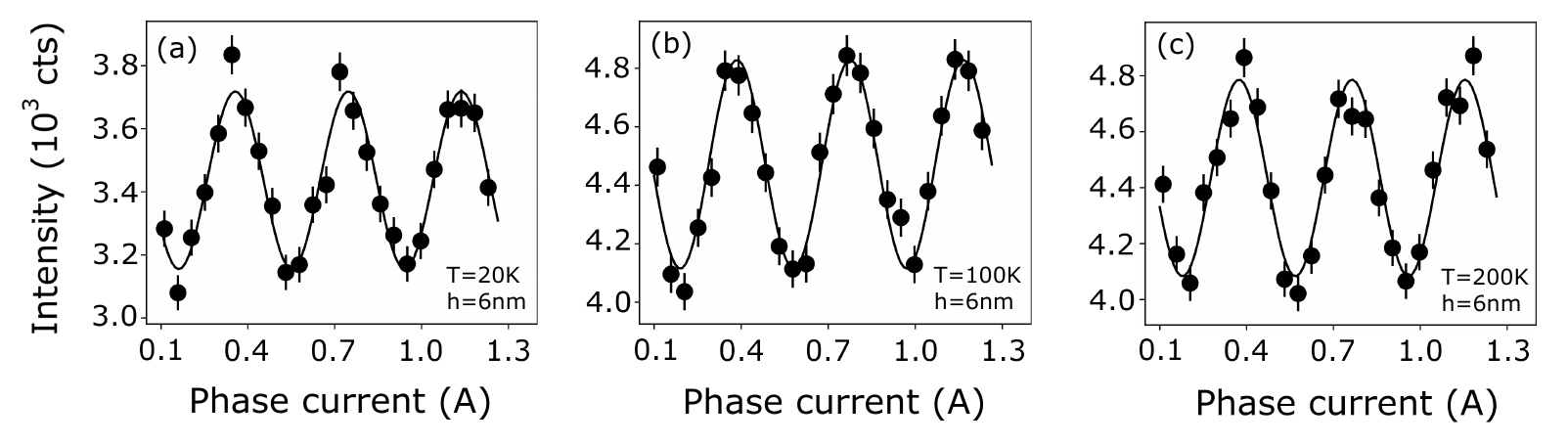}
    \caption{Echo profiles of the 6 nm honeycomb lattice at Fourier time 0.02 ns, measured at intermediate temperatures. Put together with Fig. 2b,e of the main text, the persistence of a good signal-to-background ratio from room temperature to the lowest measurement temperature (4 K) demonstrates the high quality of the data.}
    \label{fig:echoes_diff_temperature}
\end{figure}

\begin{figure}
    \includegraphics[width=0.8\textwidth]{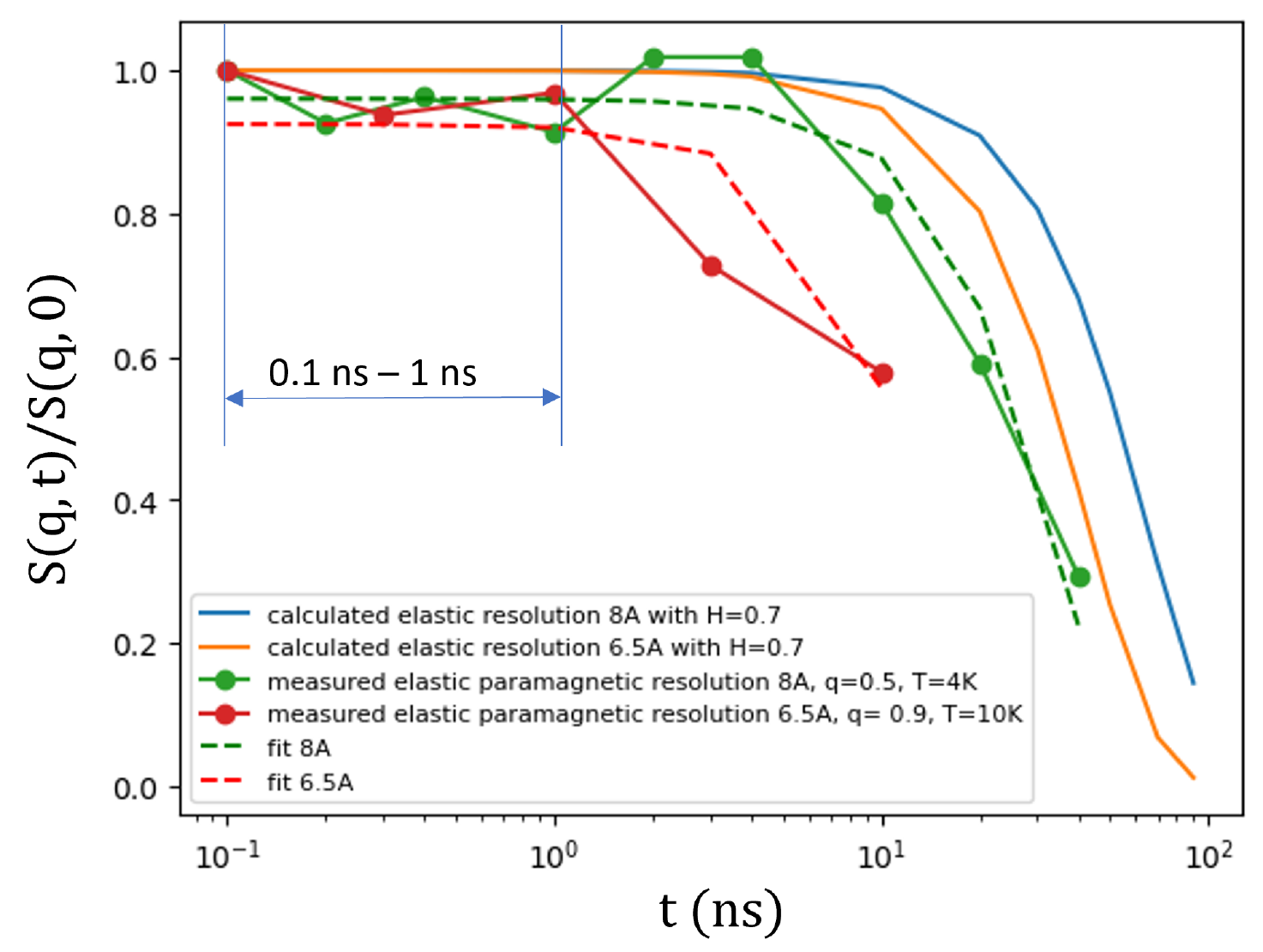}
    \caption{Measured paramagnetic resolution function at SNS-NSE and simulated elastic resolution function in soft matter regime.}
    \label{fig:NSE_resolution}
\end{figure}

\begin{figure}
    \includegraphics[width=0.49\textwidth]{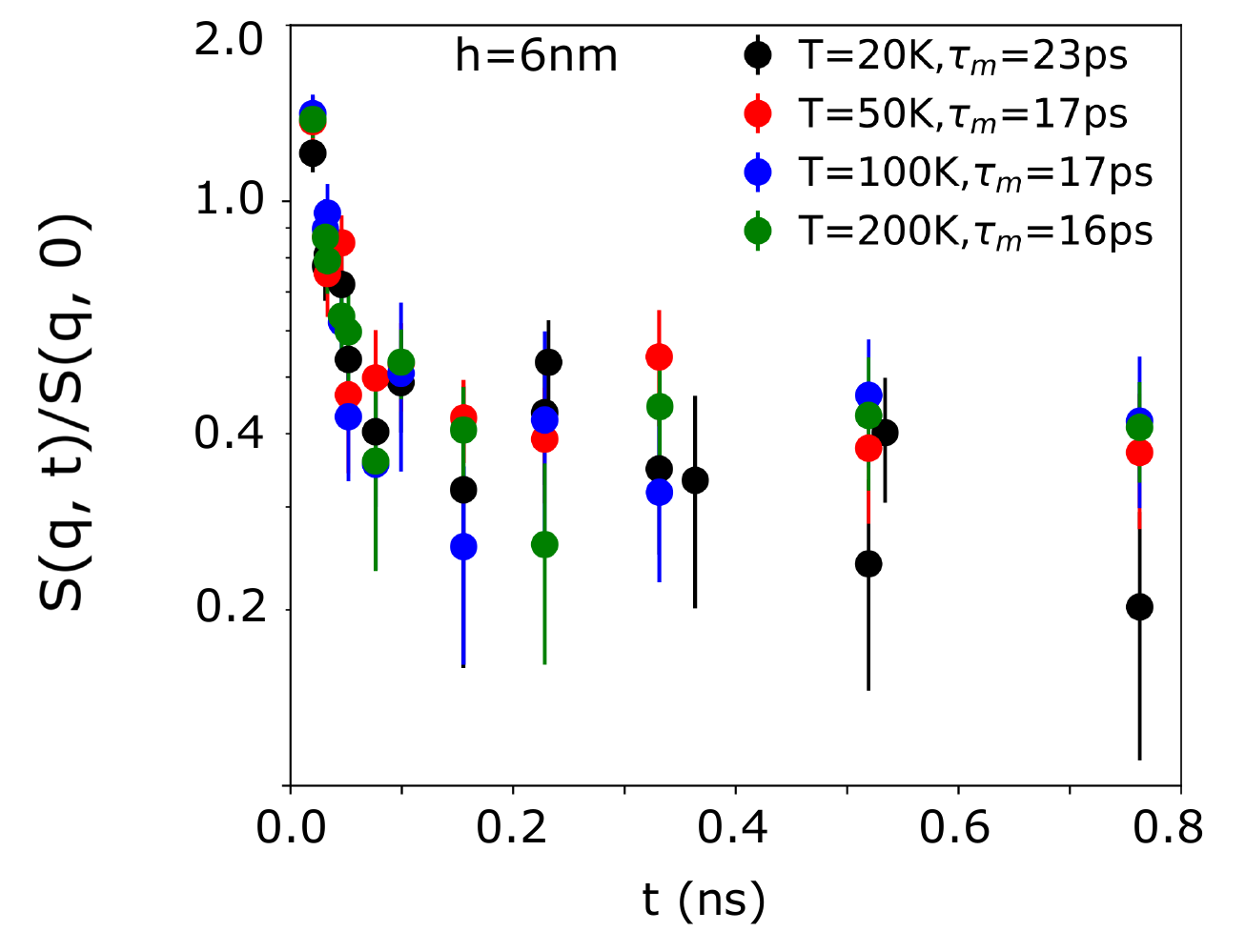}
    \includegraphics[width=0.46\textwidth]{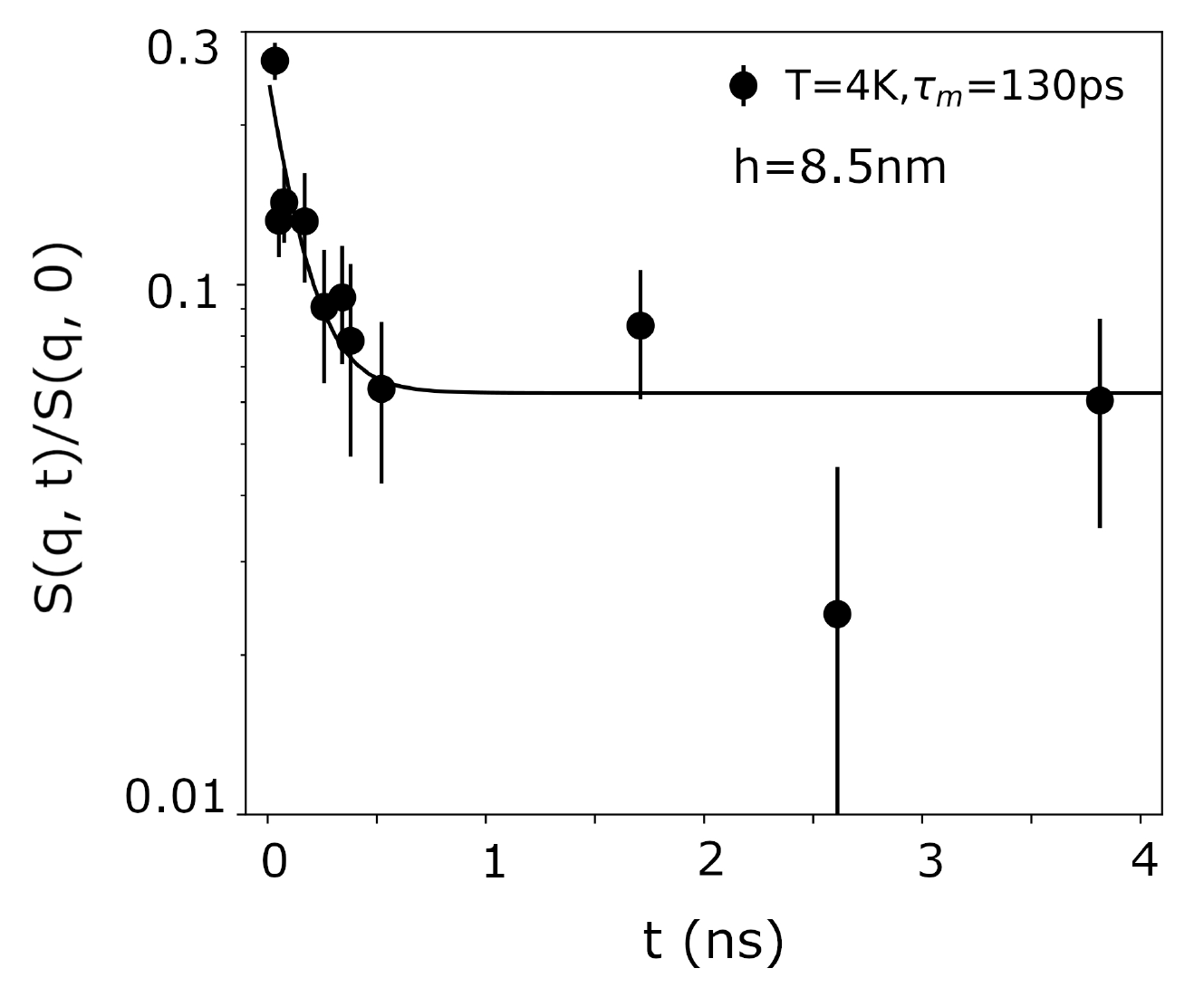}
    \caption{Left: normalized intermediate scattering function of the 6 nm honeycomb lattice at $q\sim 0.058$ \AA$^{-1}$ at intermediate temperatures, related to Fig. 3 of the main text. The relaxation time $\tau_m$ at each temperature are indicated on the graph.
    Right: normalized intermediate scattering function of the 8.5 nm honeycomb lattice at $q\sim 0.06$ \AA$^{-1}$ at $T$ = 4 K, related to Fig. 3 in the main text}
    \label{fig:suppl_sqt}
\end{figure}

Unlike experiments conducted with other spectrometers, the measured dynamic structure factor $S_{exp}(q,\omega)$ is a convolution between the true dynamic structure factor $S(q,\omega)$ and the instrument resolution $R(q,\omega)$, that is $S_{exp}(q,\omega)=S(q,\omega)\ast R(q, \omega)$. NSE directly probes the intermediate scattering function, which is the Fourier transform of the dynamics scattering function  $S(q,t)=\int{\cos(\omega t)S(q,\omega)d\omega}$, thus $S_exp (q,t)=S(q,t)R(q,t)$. Therefore, in an NSE experiment, the resolution function can be simply divided out to obtain the true intermediate scattering function. At SNS-NSE the resolution of the instrument and elastic scattering contribution is assessed by measuring a perfect elastic scattering sample, usually mounted in the same container as the sample to investigate, measured over the same q-range, tau-range, and wavelength. For soft matter measurements, solid graphite and Al$_2$O$_3$ as well as TiZr are usually used depending on the scattering angles. For paramagnetic NSE measurements we use Ho$_2$Ti$_2$O$_7$, a well-known classical spin ice material frozen below $T$ = 20 K where it exhibits no dynamics. However, for the measurement performed in this work, it was not possible to use Ho$_2$Ti$_2$O$_7$ sample as the resolution since it only scatters and produces reliable echoes at high scattering angles, while the magnetic honeycomb samples scatter in the small-angle regime. Another common practice in quasi-elastic techniques is to measure the same sample at $T \le$ 4 K where presumably all dynamics freezes. However, it is well observable in our measurements that even at $T$ = 4 K fast dynamics of the magnetic charges still exists, which is, in fact, one of the main findings of our research. Therefore, we decided not to reduce the data by elastic resolution, since, as explained, we do not have an elastic scatter with frozen dynamics within the temperature range accessible. In support of our decision is the fact that at SNS-NSE below Fourier times $t <$ 1 ns the elastic resolution is predominantly flat (linear). This means that any reduction by resolution will not affect the relaxation observed in the data but only the intensity scaling on y-axis.  To demonstrate the above we have collected two measured Ho$_2$Ti$_2$O$_7$ resolutions from previous paramagnetic measurements at $T$ = 4 K and 10 K, for 2 different wavelength 6.5 \AA~and 8 \AA. In Fig.~\ref{fig:NSE_resolution}, these experimental elastic resolution magnetic data are presented, together with their fits and simulated resolution in the soft-matter regime. One can easily observe the linear behavior in the range of $t <$ 1 ns, which is the predominant range of our measurements. Plots of $S(q,t)/S(q,0)$ at $q=0.058$ \AA$^{-1}$ at intermediate temperatures are presented in Fig.~\ref{fig:suppl_sqt}.

\section{Theoretical details}

Here we present a theoretical analysis of the ultra-small artificial magnetic honeycomb lattice, with the intention of elucidating the dominant physics at play behind the experimentally observed temperature independence of the  magnetic charge relaxation time.

As discussed in the main text, we suspect the system exhibits rapid renormalization group (RG) convergence to a minimal effective model. This intuition is motivated by the simple, consistent behavior of the magnetic charge relaxation over a broad range of temperatures; a feature that is unchanged with at least a modest change in lattice thickness. By ``rapid renormalization group convergence'' we specifically mean that small changes to a hypothetical exact theoretical description of the experimental system, such as a change in an interaction term coupling strength or a change in system size, would result in no qualitative change to the physically observable behavior of the model, due to the presence of a well isolated and strongly stable fixed point in RG space. Literature on the structurally equivalent Kagome spin ice model \cite{Chern} supports the contention that the model exhibits rapid RG convergence to the same high and low temperature phases, with or without the inclusion of long or short range interaction terms. As such we assume the system should be modeled effectively by a Hamiltonian with minimal features, e.g. a Hamiltonian that only captures the magnetic charge excitation energy and the dynamical capacity of these charges to transfer to neighboring sites.

In the following subsections, we consider two minimal effective models, one a classical description and one a quantum mechanical description, and compare the excitation relaxation time for each model to the experimental results.

\subsection{Quantum magnetic charge model}

The simplest model for a Kagome spin ice involves spins
\begin{equation}
    \vec{S}_i = \sigma_i \hat{e}_i,
    \label{eq:sp}
\end{equation}
restricted to specific site dependent directions $\hat{e}_i$ which point along three possible axes separated by $120^{\circ}$. These axes together form a honeycomb lattice structure akin to the herein studied artificial honeycomb lattices. The spins of the Kagome spin ice are coupled via a nearest-neighbor spin exchange interaction
\begin{equation}
    H_1 = - J_1 \sum_{\langle i,j \rangle} \vec{S}_i \cdot \vec{S}_j = \frac{J_1}{2} \sum_{\langle i,j \rangle} \sigma_i  \sigma_j ,
    \label{eq:h1}
\end{equation}
where $\hat{e}_i \cdot \hat{e}_i = -\frac{1}{2}$ has been used, and the notation ${\langle \dots \rangle}$ is used to indicate a sum over nearest neighbors. The system is frustrated in the case of ferromagnetic coupling, i.e. $J_1>0$, as can be seen by noting that for this sign of $J_1$ the far right side of Eq. (\ref{eq:h1}) is equivalent to an antiferromagnetic honeycomb Ising model. The Hamiltonian (\ref{eq:h1}) can be simplified further by introducing the magnetic charge at honeycomb vertex $\alpha$, $Q_{\alpha} = \pm \sum_{i \in \alpha} \sigma_i$. In terms of these operators, the Hamiltonian is equivalent to, up to a constant,
\begin{equation}
    H_1 = \frac{J_1}{4} \sum_{\alpha} Q_{\alpha}^2.
\end{equation}
Here the possible values of the magnetic charges are $Q_{\alpha} = \pm 1, \pm 3$, with the $\pm 3$ states corresponding to the localized excitations of the model.

The model described by $H_1$ is essentially classical, for the same reason that the original Ising model (without a transverse field) is classical, namely because the model lacks any non-commuting observables. This originates in the fact that, in deriving this Hamiltonian, we have neglected the important off axis components of the spins (\ref{eq:sp}). Including these off axis terms in the exchange coupling of Eq. (\ref{eq:h1}) results in nearest neighbor spin flip-flop terms (corresponding to a nearest neighbor hop of $Q_{\alpha}'s$), thus restoring the quantum nature of the model. However, instead of simply reintroducing the spin components, we consider a slightly different starting point in hopes of constructing a minimal quantum mechanical model in terms of the $Q_{\alpha}$'s. To this end, we note that the quartet nature of $Q_{\alpha}$ suggests that the simplest possible treatment of $Q_{\alpha}$ as a quantum operator would be to treat it as a $z$ component of a pseudospin 3/2 representation of SU(2). With this in mind the natural choice of non-commuting observables would be the ladder operators of the same pseudospin 3/2 representation. With a basic nearest neighbor hopping term, we arrive at the minimal quantum mechanical Hamiltonian
\begin{equation}
    H_Q = J_1 \sum_{\alpha} q_{\alpha}^2 + J_h \sum_{\langle \alpha, \beta \rangle} q_{\alpha}^+ q_{\beta}^-,
    \label{eq:QH}
\end{equation}
where we define $q_{\alpha} = \frac{1}{2} Q_{\alpha}$ in order for $q_{\alpha}$ to exhibit the traditional spin 3/2 representation eigenvalues of $\pm \frac{1}{2}$ and $\pm \frac{3}{2}$, and the ladder operators $q_{\alpha}^{\pm}$ are defined by their SU(2) commutation relations
$$
    \left[ q_{\alpha} , q_{\beta}^{\pm} \right] = \pm q_{\alpha}^{\pm} \,\delta_{\alpha\beta}, \;\;\;\;  \left[ q_{\alpha}^+ , q_{\beta}^{-} \right] = 2 q_{\alpha} \,\delta_{\alpha\beta} .
$$
In order to model excitation relaxation time using the Hamiltonian $H_Q$, we consider two separate approximation schemes, a fully integrable finite size model with 4 sites, and a dynamic Monte Carlo algorithm applied to an extended periodic boundary condition model with 200 sites.

\begin{figure}
    \includegraphics[width=0.7\textwidth]{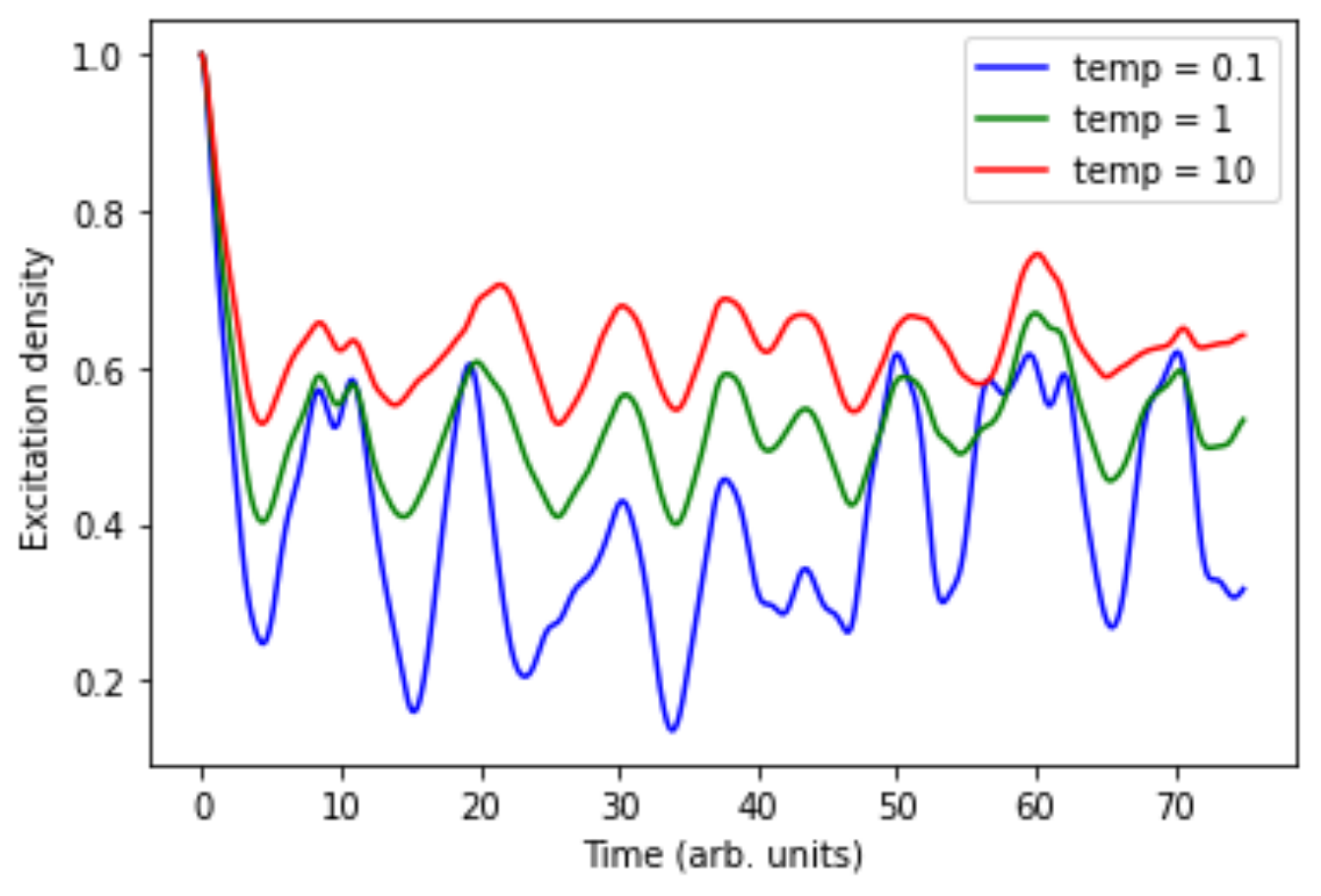}
    \caption{Exact time evolution of a four site quantum spin ice model ensemble with a central $Q=3$ excitation initial state. The curves show the density of the central charge over time for three different initial temperatures of the surrounding sites. }
    \label{fig:exact}
\end{figure}

For the first approximation, using the short range nature of the Hamiltonian, we focus on a central vertex coupled to three isolated neighbors via the flip flop term of Eq. (\ref{eq:QH}). The result is a $4^4$ dimensional Hilbert space system which we time evolve exactly. For the initial ensemble, the system is set at thermal equilibrium and projected into only states with a central $q$ eigenvalue of $q_0 = +\frac{3}{2}$. We calculate the central excitation probability $P(q_0 = +\frac{3}{2})$ as a function of time. The resulting time dependence of the $q_0 = +\frac{3}{2}$ state shows a temperature independent relaxation time, in agreement with experimental results. The main qualitative change as a function of temperature is that the large time asymptote of the time average of $P(q_0 = +\frac{3}{2})$ increases monotonically with temperature, as one would expect from ergodicity. These features can be seen in Fig.~\ref{fig:exact}.

For the second approximate treatment of (\ref{eq:QH}), we utilize the dynamic Monte Carlo algorithm as presented in Ref. 4. The system we consider has dimensions of 10 unit cells by 10 unit cells, with 2 sites per unit cell, for a total of 200 sites, under periodic boundary conditions. We time evolve the system according to the dynamic Monte Carlo algorithm through thousands of hopping processes, recording any excitations that appear in the system and computing the average lifetime of stationary excitations. In this case, it is straightforward to model the system with the addition of a potential energy barrier to magnetic charge hopping. The experimental result of temperature independent relaxation time is reproduced for the case of this barrier being set to zero, as shown in Fig. 4b. Given the fact that exchange energy is expected to be much larger than the estimated barrier height, $\sim$ 20 K (as discussed below), the magnetic charge relaxation between neighboring vertices can be considered barrier-free.

\subsection{Classical spin dynamics model}

The large number of spins per honeycomb joint in the experiment, by conventional wisdom, would suggest the possibility of modeling the system according to the purely classical treatment of a Landau-Lifshitz-Gilbert (LLG) equation of motion. Here we provide such treatment while following the same philosophy as that above of restricting ourselves to a minimalistic model. As the starting point for the above kagome spin ice was the nearest neighbor spin exchange coupling, we start with the classical analogue of this term in our LLG Hamiltonian. We further posit an anisotropy term in order to favor spins pointing along the honeycomb joint axes described by the $\hat{e}_i$ unit vectors. The result is the Hamiltonian
\begin{equation}
    H_{LLG} = - J \sum_{\langle i,j \rangle} \vec{S}_i \cdot \vec{S}_j +
    K_1 \sum_i \left(\vec{S}_i \cdot \hat{e}_i\right)^2 ,
\end{equation}
where $\vec{S}_i$ are normalized classical spin variables. Physically a term like $K_1$ arises in an effective field theory typically with dominant contributions from shape anisotropy; however this term can be used to wrap up a variety of both intrinsic and emergent anisotropic effects all into a single, phenomenological parameter. As such $K_1$'s optimal value can potentially be difficult to estimate both experimentally and theoretically. However, we have used the typical value for permalloy, $K_1 \sim -1 \times 10^4 \,\text{J/m}^3$. 

In order to model the effects of temperature, we include a time dependent stochastic field $\vec{B}_{th}$ in the LLG equations of motion, following the method of Ref. 5, which satisfies the time averaged expectation value relations
\begin{equation}
    \langle \vec{B}_{th} (t) \rangle = 0,
\end{equation}
\begin{equation}
    \langle B_{th}^i (t) B_{th}^j (t')  \rangle = \frac{2k_B T g}{M_s \gamma V \delta t} \delta_{tt'} \delta^{ij},
\end{equation}
where $\gamma$ is the gyromagnetic ratio, $g$ is the dimensionless Gilbert damping parameter, $M_s$ is the saturation magnetization, $V$ is the spatial volume of a magnetic unit cell, and $\delta t$ is the time step size of the numerical integration.

We numerically integrate a model consisting of three hexagons sharing three edges, with each edge corresponding to one spin $\vec{S}_i(t)$, for a total of 15 spins. In order to model the dynamics of a classical version of a $Q=3$ excitation, for the initial state, the three central spins are oriented outward, with the rest of the spins arranged in an energy minimizing orientation along their respective axes. For ferromagnetic coupling, this central $Q=3$ excitation state corresponds to an unstable fixed point of the LLG equations of motions. For sufficiently small temperatures, corresponding to small thermal fluctuations $\vec{B}_{th} (t)$, this model predicts an identical relaxation time for a given arrangement over a broad range of temperatures. However, we find that when the root mean square of $\vec{B}_{th} (t)$ becomes comparable to the anisotropy field $\vec{B}_{an} \propto K_1$, the relaxation time begins to decrease monotonically as a function of temperature, as shown in Fig. 4a.

The existence of this anisotropy dependent crossover allows us to estimate a thermalization crossover temperature $T_c$ for the classical model. The values used for this estimate are $\delta t = 10^{-11}$ s, $g = 0.2$, $M_s = 6.6 \times 10^5$ A/m, and $V = 264 \,\text{nm}^3$. The resulting crossover temperature estimate is on the order of $T_c \sim 20 $ K, which is much too small to be consistent with the experimental results. The substantially better agreement of the experiment with quantum modeling provides evidence for our hypothesis on the robust quantum mechanical nature of the ultra-small artificial magnetic honeycomb lattice.

We again note that an effective value of $K_1$ can be difficult to estimate due to its emergent nature and due to the fact that the shape anisotropy intrinsically depends on the geometry of the material, which in this case is highly nontrivial. In spite of this, we believe that our theoretical model supports our conclusions of the dominant physics at play.